\documentclass[prl,amsmath,amssymb,preprint,showkeys]{revtex4}
\usepackage[dvips]{graphicx}
\usepackage{amsmath}

\newcommand{\ket}[1]{\vert #1 \rangle}
\begin{document}
\title{An EIT search for Lorentz violating signals}
\author{J. P. Cotter}
\email{J.P.Cotter@sussex.ac.uk}
\author{B. T. H. Varcoe}
\email{B.Varcoe@sussex.ac.uk} 
\affiliation{Department of
Physics and Astronomy,\\ University of Sussex, \\ Falmer, Brighton \\ UK} 
\keywords{Special Relativity, Lorentz Invariance, EIT, Slow Light, Precision Spectroscopy}
\begin{abstract}
A test of Lorentz invariance has been performed using an EIT resonance as a precision frequency discriminator within an Ives-Stilwell like apparatus. We analyse the experiment within the framework of the Standard Model Extension and have made the first direct measurment of the parameter $\kappa_{tr}<1.6\times10^{-5}$.
\end{abstract}
\maketitle
The principle of Lorentz invariance has played such a central role in the foundation of modern physics that it has promoted numerous precision experiments to test the validity of its predictions. In recent times a strong motivation in the search for Lorentz violating signals has come from theories which go beyond the standard model of particle physics and allow for, if not predict, a violation from the Lorentz symmetry at some level e.g. string theory, quantum loop gravity \cite{kostelecky89}\cite{bojowald05}. Precision tests of Lorentz invariance typically take one of three forms, the Michelson-Morley test \cite{michelson87}, the Kennedy-Thorndike test \cite {kennedy32} and the Ives-Stilwell test \cite{ives38}. It is the Ives-Stilwell test we have based our experiment upon. To enhance the sensitivity of the experiment we use the modern spectroscopic technique of Electrically Induced Transparency (EIT) where a medium that is normally opaque to a laser is rendered transparent via the interaction of a second, control, laser. The same technique allows for precise manipulation of the group velocity within the medium \cite{marangos98}. The transmission window of an EIT resonance is also extremely narrow enabling precision frequency discrimination and amplifying the sensitivity of the experiment. This provides an excellent platform from which to look for new physics \cite{varcoe06}.

There are a number of test theories capable of analysing experiments sensitive to a deviation from the Lorentz symmetry which offer useful comparisons between different experiments \cite{robertson49}\cite{mansouri77}. However, it is hard to gain insight into the relevant underlying physical processes and therefore deomstrate the subtlty of different approaches. Recently, Kostelecky and co-workers have parameterised violations of Lorentz invariance within an extension of the Standard Model of particle physics \cite{colladay97}\cite{colladay98}. The Standard Model Extension (SME), as it is known, allows for small general CPT and Lorentz violating terms constructed from known fields. The fermion sector of this general theory has received much attention, and precise bounds have been placed on its predictions \cite{kostelecky02}. In this paper we therefore ignore this sector and look at terms concerning only the QED sector of the SME \cite{colladay97}\cite{colladay98} with a resulting lagrangian density,
\begin{eqnarray}
{\cal L}=&-&\frac{1}{4}F_{\mu \nu}F^{\mu \nu}+\frac{1}{2}(k_{AF})^{\kappa}\epsilon_{\kappa \lambda \mu \nu}A^{\lambda}F^{\mu \nu}\nonumber\\
&-&\frac{1}{4}(k_{F})_{\kappa \lambda \mu \nu}F^{\kappa \lambda}F_{\mu \nu}~,\label{lagrangian}
\end{eqnarray}
where $F_{\mu \nu}$ is the normal Maxwell field tensor, $A^{\lambda}$ is the vector potential and the two co-efficients $(k_{AF})^{\kappa}$ and $(k_{F})_{\kappa \lambda \mu \nu}$ describe Lorentz violating terms which are CPT odd and CPT even respectively. Experiments \cite{kostelecky02} have placed stringent bounds on $(\kappa_{AF})^{\kappa}$, and for our purposes we have assumed it to be zero, $(\kappa_{AF})^{\kappa}=0$. To understand the relationship between our experiment and the SME we draw upon a useful analogy between photon propagation in this modified Lorentz violating form of QED and classical homogeneous anisotropic media \cite{colladay97}\cite{colladay98}:
\begin{equation}
\left(\begin{array}{c}\bf{D}\\\bf{H}\end{array}\right)=\left(\begin{array}{cc}1+\kappa_{DE}&\kappa_{DB}\\\kappa_{HE}&1+\kappa_{HB}\end{array}\right)\left(\begin{array}{c}\bf{E}\\\bf{B}\end{array}\right)\label{kosteleckymatrix}
\end{equation}
The 3x3 matrices $\kappa_{DE}$, $\kappa_{DB}$, $\kappa_{HD}$ and $\kappa_{HB}$ contain 19 independent parameters, considered to be constant in both space and time, which describe a violation from the Lorentz symmetry. A detailed description of these matrices in terms of the CPT even co-efficient $(k_{F})_{\kappa \lambda \mu \nu}$ from equation (\ref{lagrangian}) can be found in Ref \cite{kostelecky02}.
Each $\kappa$-tensor from equation (\ref{kosteleckymatrix}) can be re-expressed in terms of quantities accessable directly via experimentation:
\begin{eqnarray}
\kappa_{jk}^{DE}&=&(\kappa^{e+}+\kappa^{e-})_{jk}+\kappa_{tr}\delta_{jk},\nonumber\\
\kappa_{jk}^{HB}&=&(\kappa^{e+}-\kappa^{e-})_{jk}-\kappa_{tr}\delta_{jk},\nonumber\\
\kappa_{jk}^{DB}&=&(\kappa^{o+}+\kappa^{o-})_{jk},\nonumber\\
\kappa_{jk}^{HE}&=&(\kappa^{o+}-\kappa^{o-})_{jk}.
\end{eqnarray} 
Astrophysical observations have constrained the ten parameters in $\kappa_{e+}$ and $\kappa_{o-}$  to a part in $10^{32}$ \cite{kostelecky02}. The remaining nine parameters, five in $\kappa_{e-}$, three in $\kappa_{o+}$, and a scalar $\kappa_{tr}$, can be accessed via precision table top experiments \cite{lipa03}\cite{muller03}\cite{schiller05}\cite{tobar05}\cite{saathoff03}\cite{herrmann05}\cite{stanwix06}. Upper bounds of $10^{-15}$ and $10^{-11}$  have been placed on terms in the tensors $\kappa_{e-}$ and $\kappa_{o+}$ respectively using precision Michelson-Morley experiments. While $\kappa_{tr}$, related to the effective permeability, permittivity and therefore absolute velocity of light in the vacuum, has been constrained to a part in $10^{5}$ \cite{saathoff03} analysis in \cite{tobar05}. The relatively poor knowledge of $\kappa_{tr}$ compared with the other SME parameters allows us to make the simplification that $\kappa_{e-}$, $\kappa_{e+}$, $\kappa_{o-}$ and $\kappa_{o+}$ are zero.  Allowing us to reduce $\kappa_{jk}^{DB}$,$\kappa_{jk}^{HE}$,$\kappa_{jk}^{DE}$ and $\kappa_{jk}^{HB}$ to
\begin{eqnarray}
\kappa_{jk}^{DB}&=&\kappa_{jk}^{HE}=0\label{k1},\\
\kappa_{jk}^{DE}&=&\kappa_{tr}\delta_{jk}\label{k2},\\
\kappa_{jk}^{HB}&=&-\kappa_{tr}\delta_{jk}\label{k3}.
\end{eqnarray}
At this point we can make further use of the analogy between light propagation ({\it in vacuo}) in the SME and in classical anisotropic media by introducing the refractive index vector $n_{i}$. From Maxwell's equations in anisotropic media we obtain an expression for the electric displacement field in terms of the refractive index vector and electric field only:
\begin{eqnarray}
D_{j}&=&\epsilon_{jkl}n^{k}\epsilon^{lmn}E_{m}n_{n}\nonumber\\
&=&n^{2}E_{j}-n^{i}E_{i}n_{j}\label{refind1}.
\end{eqnarray}
By comparing this to the explicit form of the electric displacement field within the SME,  $D_{j}=(1+\kappa^{DE})_{jk}E^{k}+ \kappa^{DB}_{jk}B^k$, and making use of equations (\ref{k1}) and (\ref{k2}) we arrive at three linear homogeneous equations involving only the refractive index vector and $\kappa_{tr}$ for the three components of the electric field vector,
\begin{equation}
(n^{2}\delta_{jk}-n_{j}n_{k}-(1+\kappa_{tr})\delta_{jk})E^{k}\label{det}=0.
\end{equation}
Non-trivial solutions of this equation result in an expression for the magnitude of the refractive index in terms of $\kappa_{tr}$ alone \cite{lan_ECM}:
\begin{equation}
n=\sqrt{1+\kappa_{tr}}.
\end{equation}

The $\kappa$-tensors are frame dependent unless the observer is in a cosmological frame; i.e a frame defined by the Cosmic Microwave Background, or any frame moving inertially with respect to it. The frame dependency of $\kappa_{tr}$ will induce a frame dependency in the phase velocity of light. Ives-Stilwell experiments \cite{ives38}, and their modern counterparts, are sensitive to a deviation between the factor $c$, appearing in the Lorentz time dilation $\gamma=1/\sqrt{1-v^{2}/c^{2}}$, and the phase velocity of light $c'$ appearing in the Doppler shift formula \cite{varcoe06}. By comparing the Doppler shifted frequencies of two counter propagating lasers to a frequency reference, the predictions of time dilation can be tested and a measurement of $\kappa_{tr}$ made.
\begin{figure}[!ht]
\begin{center}
\scalebox{0.39}{\includegraphics{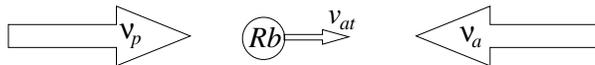}}
\caption{Modernised Ives-Stilwell set-up: An effective atomic beam of velocity $v_{at}$ is used to measure 
the Doppler shifted frequencies $\nu_{p}$ and $\nu_{a}$ of two counter propagating lasers 
moving parallel and anti-parallel to an atomic beam.}
\end{center}
\end{figure}

It is conventional to analyse and compare experiments concerning parameters from the SME in a co-ordinate system with the sun centred at the origin. Such a frame ignores the rotational motion of the sun. Thus the SME shows that in the presence of a Lorentz violating effect there will be a frequency difference, $\Delta\nu=|\nu_{a}-\nu_{p}|$,  between the two counter-propagating lasers of an Ives-Stilwell experiment. In the sun centered frame $\Delta\nu$ will be dependent on the sidereal modulations in the earth's velocity due to its rotation, and the annual modulations due to it's orbital motion about the sun. Note that because the SME has been constructed so as to maintain invariance under Lorentz transformations between inertial frames we are still able to transform refractive indices according to the conventional laws of special relativity, and after some calculation we arrive at a first order approximation for the modified refractive index magnitude in the sun centred frame, $n'=1+\kappa_{tr}\left(1/2-v_{lab\parallel}/c\right)$. Here $v_{lab\parallel}$ is the velocity of the laboratory with respect to this frame moving along a line parallel to the experimental axis. Working in a time frame much less than a year we can look for periodic signals containing only a 24-sidereal-hour signal. Therefore, for non-vanishing values of $\kappa_{tr}$ we arrive at an expression for the additional frequency splitting caused by a violation of Lorentz invariance as observed in the sun centred frame,
\begin{equation}
\Delta \nu_{mod}=2\nu_{0}\beta_{at}\beta_{lab}\kappa_{tr}\cos(\omega t + \phi),
\end{equation}
where $\Delta \nu_{mod}$ is the periodic difference frequency between our two lasers, $\nu_{0}$ is the rest frame frequency of the laser, $\beta_{at}\sim4\times10^{-6}$, $\beta_{lab}\sim1\times10^{-3}$, $\omega$ is the sidereal angular frequency of the earth, t is time in the lab frame and $\phi$ is a phase factor dependant upon the position of the earth in its orbit.

We used the Sagnac interferometer of Jundt et al \cite{jundt03} as the basis for our experiment. An effective beam of $Rb^{85}$ atoms, orientated east-west, is used as a frequency reference to measure the Doppler shifted frequencies of two lasers moving in parallel, $\nu_{p}$, and anti-parallel, $\nu_{a}$, directions with respect to the atomic beam. The beauty of this configuration is its inherent stability against small changes in cavity length because both arms of the interferometer use the same optics. Thus the two counter propagating lasers necessary for an Ives-Stilwell experiment are formed from the two interfering beams that make up our interferometer. This ensures both lasers have the same rest frame frequency. Using a dual isotope, rubidium cell with a buffer gas of neon at a pressure of 30Torr we form a EIT resonance with a measured linewidth of $\sim300$ Hz. The diode laser ($\lambda=795$ nm) has a linewidth of $\sim1$ MHz and is tuned so that its rest frame frequency is halfway between the $\ket{F=2} \leftrightarrow \ket{F'=2}$ and $\ket{F=3}\leftrightarrow \ket{F'=2,}$ hyperfine transition frequencies of the $5^{2}S_{1/2}\leftrightarrow5^{2}P_{1/2}$ energy levels in $^{85}Rb$. Atoms within the cell that are moving along the laser axis with a critical speed of $v_{at}\sim\pm1200$ ms$^{-1}$ relative to the lab are Doppler shifted into resonance with one of these transitions. If the Lorentz symmetry is broken the lasers will acquire a directional frequency shift. The result being that the joint resonance condition between the atom and counter propagating light fields is no longer satisfied.
\begin{figure}[!ht]
\begin{center}
\scalebox{0.3}{\includegraphics{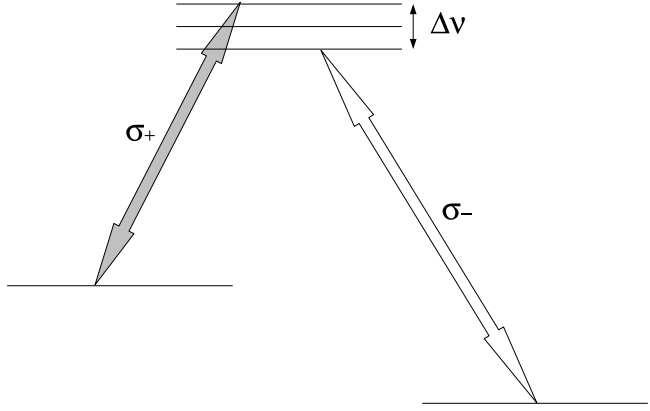}}
\caption{Three level $\Lambda$-system in the case of Lorentz violation. $\Delta \nu$ measures the frequency splitting caused by a violation of Lorentz invariance. In this figure the grey arrow corresponds to the intense pump field and the white arrow the weaker probe field. However, the symmetry of the experiment means that there is an equivilent situation for atoms moving in the opposite direction with the pump and probe beam frequencies interchanged}
\end{center}
\end{figure}

The detuning between $\ket{F=2} \leftrightarrow \ket{F'=2}$ and $\ket{F=3}\leftrightarrow \ket{F'=2}$ magnetic substates is controlled using an external magnetic field. A concentric solenoid is positioned between the cell and double layered mu-metal shielding. Scanning the magnetic field and keeping the laser frequency constant allows us to determine the difference frequency, $\Delta\nu$, between the light beams. The spatial symmetry of the experiment provides us with two simultaneous measurements. Those atoms moving in opposite directions observe the pump and probe beam frequencies interchanged. For non-zero values of $\kappa_{tr}$ we expect to see two EIT dips with a separation proportional to the magnitude of $\kappa_{tr}$. Other experiments have shown that any such violation will be extremely small \cite{saathoff03}\cite{tobar05}, even with a narrow transmission window of $~300$Hz it will not be possible to resolve each of the two peaks individually. In our data analysis we therefore assume that the observed dip in the absorption of the probe beam is the convolution of two unresolved dips using a fit  to extract the separation between them. It is expected that any observed signal will take the form of a 24-sidereal-hour cosine. However, because we are sensitive only to the magnitude of the beat frequency, we have no means of distinguishing beams after detection, we therefore look for signals with an angular frequency $2\omega$ only. The atomic velocity distribution within the cell is removed by the use of counter propagating lasers. A detailed description of the physics of this interaction and the nature of Doppler free spectroscopy will be included in a future publication.

\begin{figure}[ht]
\scalebox{0.3}{\includegraphics{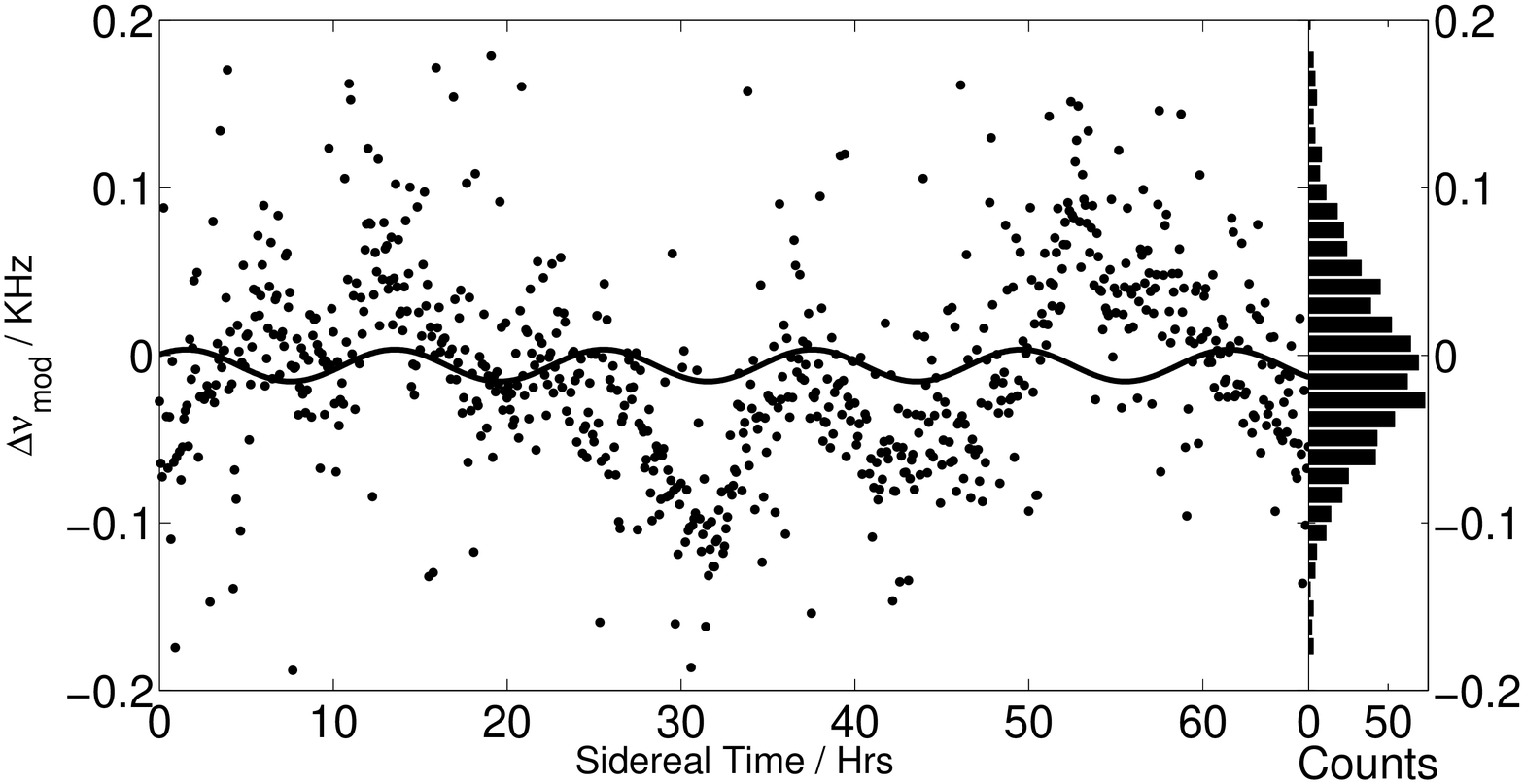}}
\caption{Experimental data and the fitted 12-sidereal hour modulation. A constant offset has been subtracted. The amplitude of the fit is the only free parameter: The phase is fixed by the position of the earth in its orbit at the time the experiment was started. The histogram on the right hand side shows the distribution of residuals.}
\label{data}
\end{figure}

Data is collated by measuring the difference frequency $\Delta\nu$ at 5 minute intervals. The data presented in Fig.\ref{data} has been extracted from a run performed during April 2005. The fitted separation between two unresolved dips is displayed as a function of time over a 72-hour period. In the present paper we are only concerned with a relative change in $\Delta\nu$ and not an absolute measurement of frequency shift which is more sensitive to background magnetic fields; a constant offset has been removed from the data. In order to look for modulations in the data we have used a least squares technique to fit an amplitude to a 12-sidereal-hour cosine with a phase fixed by the position of the earth in its orbit and the time of day.
The amplitude of the fitted cosine is $11 \pm 51$ Hz (90\% confidence). Thus the largest signal concealed within our data is,
\begin{equation}
\kappa_{tr}<1.6\times10^{-5},
\end{equation}
to within a 90\% confidence interval. We believe this to be the best direct measurement of $\kappa_{tr}$ to date. This upper limit is comparable to the $\kappa_{tr}\lesssim10^{-5}$ result of Tobar et al \cite{tobar05}  extracted from the measurement of Saathoff \cite{saathoff03} via analysis within the SME.

There are two paths to increase the sensitivty of the experiment: removing residual magnetic fields, and increasing the atomic velocity which is linearly proportional to the predicted frequency splitting. Magnetic fields present within the shielding cause frequency splittings, and have been shown to contain components with a modulation frequency comparable to that of our 12-sidereal hour signal that will contribute a systematic effect. The most likely cause of this being the temperature dependant permeabillity of the mu-metal. For future experiments we propose moving to a less magnetically susceptible $\ket{m_{f}=0}\rightarrow\ket{m_{f}=0}$ transition using linearly polarised light. Considering such transitions are only subject to the much smaller second order Zeeman shift this will greatly reduce the effect of residual fields. While increasing the atomic velocity will improve the sensitivity of the experiment an EIT resonance must be found with suitable hyperfine transition such that $v_{at}=\Delta\nu_{hyp} c /2\nu_{0}$. Further room for improvement where large gains in sensitivity can be achieved is in analysis of a constant ofset in the frequency differences. These needed to be discarded in the current experiment as residual magnetic fields also give rise to similar shifts. Removing the effect of this systematic shift permits this measurement and can provide up to three orders of magnitude improvement in the resolution.

\acknowledgments
\subsection{Acknowledgements}
This work is supported by PPARC's Particle Physics Peer Review Panel. Grateful thanks must also go to Peter Blythe, Claudia Eberlein and Robert Smith for useful discussion. 
\bibliographystyle{apsrev}
\bibliography{Bibliography}
\raggedright
\raggedbottom
\end{document}